# Detection of circular permutations by Protein Language Models

*Yue Hu[1#*], Bin Huang[2#*](co-first and corresponding author), Chun Zi Zang[3]*

*[1]College of Bioengineering, Qilu University of Technology (Shandong Academy of Sciences), Jinan 250300, Shandong, China*

*[2] School of Life Sciences, Yunnan Normal University, Kunming 650500, Yunnan, China*

*[3]Kyiv College, Qilu University of Technology (Shandong Academy of Sciences), Jinan 250300, Shandong, China*

*#Co-first author*

**Corresponding Author：*

*Yue Hu, E-mail: huyue@qlu.edu.cn*

*Bin Huang, E-mail: huangbin@ynnu.edu.cn*

***Protein circular permutations are crucial for understanding protein evolution and functionality. Traditional detection methods, sequence-based or structure-based, struggle with accuracy and computational efficiency, the latter also limited by treating proteins as rigid bodies. The plmCP method, utilizing a protein language model, not only speeds up the detection process but also enhances the accuracy of identifying circular permutations, contributing significantly to protein research and engineering by acknowledging structural flexibility.***

Circular permutation in proteins[1-4] is a fascinating phenomenon that reflects the versatility and adaptability of these biological macromolecules. A circular permutation occurs when the order of amino acids in a protein's sequence is rearranged, such that the N-terminus and C-terminus are shifted (Figure 1(A)). Despite this rearrangement, the overall three-dimensional shape and function of the protein often remain conserved (Figure 1(A)). In proteins, they may occur through mechanisms such as gene duplication followed by the loss of redundant sections (which is also the principle to detect circular permutation by sequence alignment), or through fission and fusion events where partial proteins combine to form a new polypeptide. And the third mechanism maybe occurred by the cyclization and cleavage of circRNA or circProtein (Figure 1(B)). These permutations can lead to proteins with altered catalytic activities, increased thermostability, or novel functionalities[5, 6]. The ability to detect and study such permutations enhances our understanding of protein evolution and enables the development of proteins with improved or novel characteristics.

Detecting circular permutations in proteins is challenging due to the non-linear nature of the rearrangement[1, 4, 7, 8]. Algorithms for detecting circular permutations in proteins can be divided into sequence-based and structure-based methods. SeqCP[9] is a sequence-based algorithm for searching circularly permuted proteins. It analyses normal and duplicated sequence alignments to identify circularly permuted regions (followed the first occurrence mechanisms described above). It should be noted that this algorithm has taken into account all possible circular permutation, regardless of the mechanism of occurrence. It also has distinct advantages, one is fast calculation speed, and the other is that it does not need to know the structure of proteins. CE-CP[7], TMalign[10, 11] and CPSARST[12] are three typical structure-based methods for detecting circular permutations. Although CPSARST is a structure-based method, in reality, it represents the information of the structure as a string, and when comparing the structure, it is similar to sequence alignment, which not only utilizes the accuracy of structure alignment but also the speed of sequence alignment.

The circular permutation detection method based on sequence dependencies may be fast, but it may sacrifice some accuracy, especially for two remote evolutional proteins (with low sequence similarity). The detection of such permutations is traditionally reliant on structural alignment methods, which, while accurate, are computationally intensive and require known three-dimensional structures. In summary, traditional detection methods, structure-based or sequence-based, or struggle with accuracy and computational efficiency, the latter also limited by treating proteins as rigid bodies.

Protein language models can help us extract structural information from the co-evolution information of sequences in implicit way. Detecting circular permutations in proteins is enhanced by Protein Language Models (PLMs), which, inspired by plmAlign[13], DeepBlast[14], pLM-BLAST[15],DEDAL[16], use the inner product of amino acid vectors to construct a density matrix and a subsequent scoring matrix. This matrix is then utilized by the Smith-Waterman algorithm to find the best local alignment between sequences. By comparing the alignment of a query sequence with the target sequence and its duplicate with the target, and applying simple criteria, we can determine the presence of a circular permutation. Using the above ideas, we have constructed plmCP (protein language models for the detection of circular permutations) algorithm (Figure 1(C)).

First, we utilize two remote evolutional proteins 3CNA and 2PEL (sequence identity about 15%) as the case to verify the ability to detect the remote circular permutation by plmCP. The TM-align tool to compare those two proteins using two methods: one with circular permutation (TM-score=0.90942) and the other without (TM-score=0.48028). These methods yield significantly different results. With circular permutation, the two proteins superpose perfectly, indicating that the TM-align tool can accurately detect this circular permutation pair (Figure 1(A)). Next, we employ plmCP to analyse the same proteins, obtaining the CP1 value (0.945) is approximately 1 and the CP2 value (0.000655) is close to 0 (see the definition of parameters in Method), suggesting that these proteins may indeed undergo circular permutation. Finally, when comparing the alignments from TM-align (with CP parameter) and plmCP, we observe obvious similarities between the two methods. (Figure 1(A)).

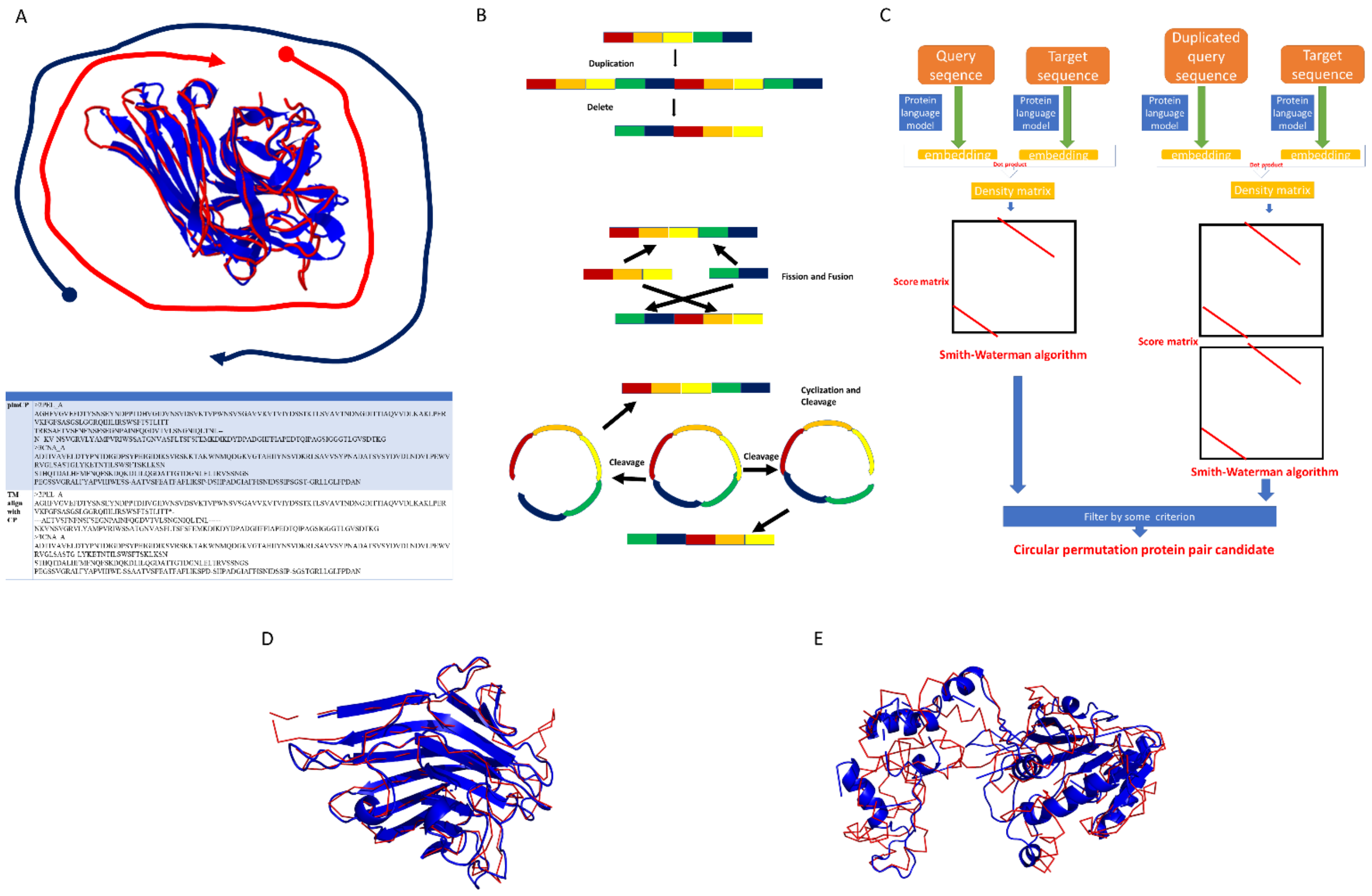

**Figure 1. (A) Superposition of two circular permutation proteins (2PEL in blue and 3CNA in red) and the alignment results of TMalign and plmCP. (B) The possible mechanisms of circular permutation occur. (C) The principal diagram of the plmCP. (D) Superposition of two circular permutation proteins (1NLS in blue and 2BQP in red). (E) Superposition of two circular permutation proteins (1NW5 in blue and 2ADM in red).**

We then utilize another protein pairwise comparison tool (CE, CE-CP, FATCAT[17](rigid) and FATCAT(flexible)) to superpose the proteins and compute the TM-score, which give TM-score of 0.22, 0.32, 0.15, 0.09, respectively. CPSARST is able to identify the relationship when the query is 2PEL, but it does not find any relation when searching with 3CNA. The sequence-based method, seqCP, fails to detect any relationship between the proteins. These tools perform poorly with this protein pairs.

To further validate our approach, we employed a challenging database (the RIPC database[18]) that specifically contains protein structures with circular permutations. This database also provides reference information for sequence alignment[19]. First, we focused the performance of plmCP, TMalign, and ICARUS[19]. In the case of 1NLS and 2BQP (Figure 1(D)), we demonstrated the consistency among the three programs. In another case of 1NW5 and 2ADM (Figure 1(E)), our program plmCP performed exceptionally well, outperforming both TMalign (which has a very low TM-score) and ICARUS. Those protein pairs have a context of circular permutation and insertion structural variations, which will make it hard to align. It's worth noting that ICARUS incurs significant computational cost. Compared with reference of structures contained circular permutation type in RIPC database, the plmCP archived 66.47% accurately aligned positions, while the structure methods range from 27.85% to 76.05%. This proves the effectiveness of plmCP software in detecting circular permutations. Beyond analysing the plmCP's performance in circular permutation detection, we also compared other types of structural variations in the database to the reference results. Compared with reference, the plmCP archived 78.3% accurately aligned positions, while the structure methods range from 56.6% to 81.2%. The results demonstrate that our method can perform comparative detection for various structural variations, confirming its universality and expanding its applicability. Overall, we proposed a sequence-based, non-linear, flexible protein alignment tool.

## Method

The plmCP model is designed to identify circular permutations in protein sequences by leveraging the power of protein language models and the principles of pairwise sequence alignment algorithm. The method inspired by plmAlign and CPSARST. The plmCP (Figure 1 (C)) employs state-of-the-art protein language models such as ESM-1b to generate per-residue embeddings. These embeddings represent the individual amino acids in the context of the entire protein sequence, capturing the intricate relationships and structural features that might not be evident from the primary sequence alone. The plmCP aligns the query sequence with the target sequence and vice versa. Additionally, it aligns a duplicate of the query sequence (or target sequence) with the target (or query). Using the inner product of amino acid vectors to construct a density matrix (Equation (1)) and a subsequent scoring matrix (Equation (2)). This matrix is then utilized by the Smith-Waterman algorithm to find the best local alignment between sequences. Importantly, it changes to the classical backtracking algorithm in Simth-Waterman algorithm (Equation (3)) compared with plmAlign. By comparing the results of these two alignments (the alignment score of Smith-Waterman algorithm, the aligned length, and some parameters defined below), plmCP can detect circular permutations, which may arise from gene duplication followed by deletion events. The plmCP software based on the codes of plmAlign and combined with CPSART algorithm to detect circular permutations of proteins. The source code can be obtained in the supplementary and can be obtained from https://github.com/YueHuLab/plmCP.

$$Density = {E_{query}}^{T} E_{target} \tag{1}$$

$$score(i,j) = max \begin{cases} score(i-1,j) \\ score(i,j-1) \\ score(i-1,j-1) + density(i,j) \\ 0 \end{cases} \tag{2}$$

$$\mathrm{trackback}(i, j) = \begin{cases} 3,\ if\ score(i,j) == score(i-1,j) \\ 2,\ \ if\ score(i,j) == score(i,j-1) \\ 1,\ if\ score(i,j) == score(i-1,j-1) + density(i,j) \\ 0,\ if\ score(i,j) == 0 \end{cases} \tag{3}$$

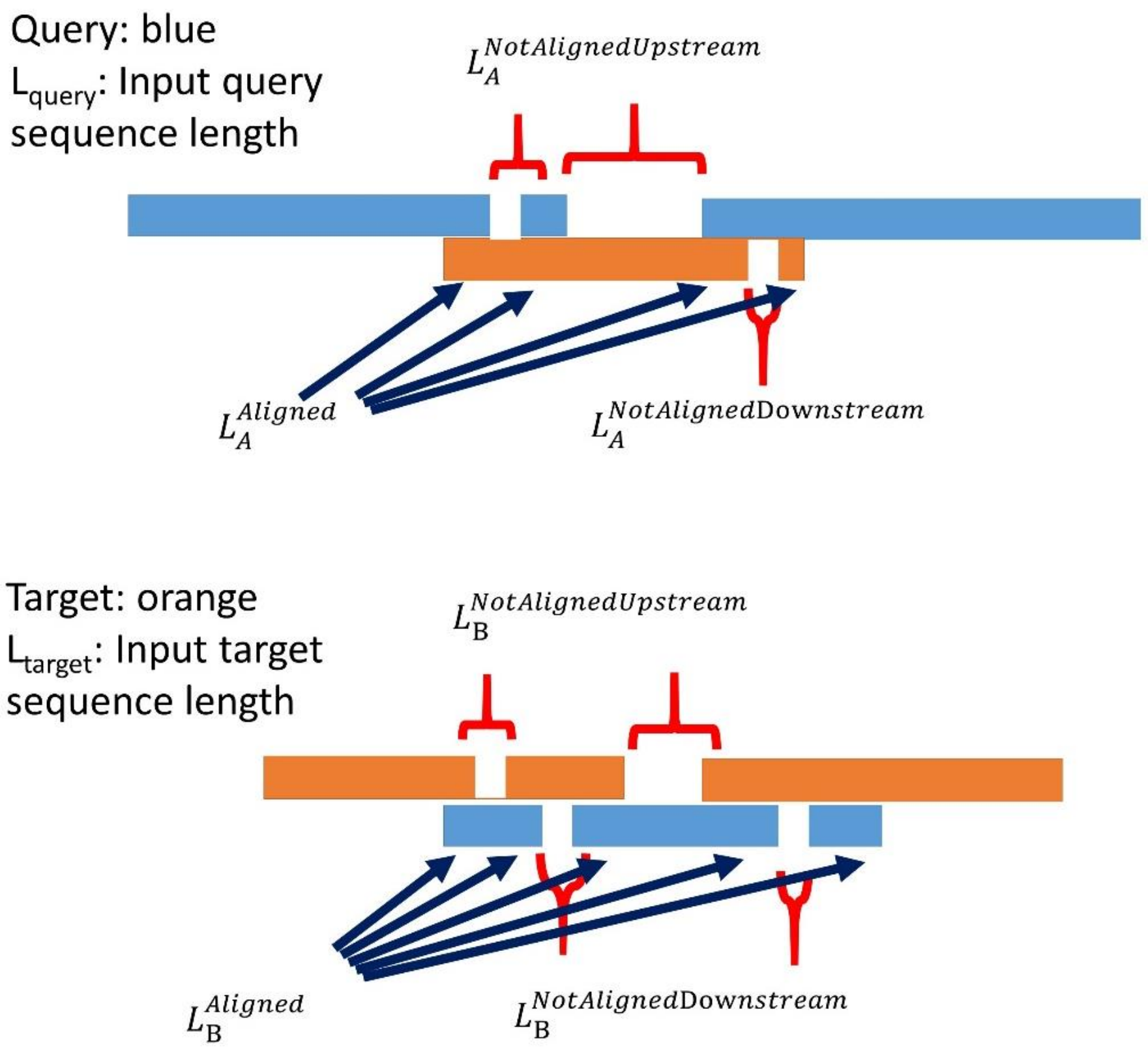

**Figure 2. Schematic diagram of parameter definition.**

As shown in Figure 2, we use some simple parameters to perform a rough screening of circular permutation. In an ideal state, except for the end positions, two circular permutation proteins should be completely overlapping. In this case, the value of CP1 (Equation (4)) should be equal to 1, and the value of CP2 (Equation (5)) should be equal to 0. In the meantime, we should know if one protein aligns with itself, the value of CP1 will also be equal to 1, and the value of CP2 will also be equal to 0. Indeed, the protein may by circular permutated from itself. In this circumstance, the alignment will from the beginning. So, we can check the alignment to exclude this circumstance. In order to analyse the alignment of each sequence individually, we defined 6 parameters (Equation (6-11)). Equation (6) and (9) reflect the aligned proportion compared to target length or query length. Equation (7) and (10) reflect the insertion of upstream sequence compared target length or query length. Equation (8) and (11) reflect the insertion of upstream sequence compared target length or query length.

$$CP1 = \frac{L_A^{Aligned} * L_B^{Aligned}}{L_{target} * L_{query}} \quad (4)$$

$$CP2 = \frac{(L_{target} - L_A^{Aligned}) * (L_{query} - L_B^{Aligned})}{L_{target} * L_{query}} \quad (5)$$

$$CP_{check_A} = \frac{L_A^{Aligned}}{L_{target}} \quad (6)$$

$$CP_{check_{AN1}} = \frac{L_A^{NotAlignedUpstream}}{L_{target}} \quad (7)$$

$$CP_{check_{AN2}} = \frac{L_A^{NotAlignedDownstream}}{L_{target}} \quad (8)$$

$$CP_{check_B} = \frac{L_B^{Aligned}}{L_{query}} \quad (9)$$

$$CP_{check_{BN1}} = \frac{L_B^{NotAlignedUpstream}}{L_{query}} \quad (10)$$

$$CP_{check_{BN2}} = \frac{L_B^{NotAlignedDownStream}}{L_{query}} \quad (11)$$

In the evaluation of the PLMCP model, we utilized the Protein Data Bank (PDB) entries 3CNA (concanavalin) and 2PEL(peanut lectin) as the test case to demonstrate the efficacy of our method. The plmCP model's performance was benchmarked against established methods such as TMalign, CE-CP, CE, SeqCP, FATCAT and CPSARST. The comparison was based on two key metrics: the TM-score, which assesses the structural similarity between protein pairs; and the sequence alignment results, which measure and demonstrate the differences in specific comparisons. To further validate our approach, we employed a challenging database (the RIPC database) that specifically contains protein structures with circular permutations. This database also provides sequence alignment information for reference. We compared plmCP with TMalign and ICARUS in the different types of structural variation in this database.

## Acknowledgements

This work was supported by the National Natural Science Foundation of China (Grant No. 32301061 to B.H.) and was supported by the Yunnan Fundamental Research Projects (Grant No. 202401CF070034 to B.H.) and was supported by Shandong Provincial Natural Science

Foundation Committee (Grant No. ZR2016HB54 Y.H.) and was supported by Shandong Provincial Key Laboratory of Microbial Engineering (SME).

## Author contributions

Hu Yue and Huang Bin secured funding and designed the algorithm, who are also guarantors of this work and, as such, had full access to all of the data in the study and takes responsibility for the integrity of the data and the accuracy of the data analysis; Hu Yue, Huang Bin, Zang ChunZi performed computation and analyzed the data.

## Competing interests

All authors declare no competing interests.